\documentclass{article}

\usepackage[english]{babel}
\usepackage[ansinew]{inputenc}

\usepackage{amsfonts}
\usepackage{amssymb}
\usepackage{subfigure}

\usepackage{psfrag,graphicx}
\usepackage{natbib}

\newcommand{\app}[3]{Acta Phys. Pol. \textbf{#1} ({#2}) {#3}}
\newcommand{\nucl}[3]{Nucl. Phys. \textbf{#1} ({#2}) {#3}}

\newcommand{\prd}[3]{Phys. Rev. \textbf{D{#1}} ({#2}) {#3}}
\newcommand{\prev}[3]{Phys. Rev. \textbf{{#1}} ({#2}) {#3}}
\newcommand{\plb}[3]{Phys. Lett. \textbf{B{#1}} ({#2}) {#3}}
\newcommand{\prep}[3]{Phys. Rep. \textbf{{#1}} ({#2}) {#3}}

\newcommand{\nipop}{$N_{ipop} ~$}
\newcommand{\npop}{$N_{pop} ~$}
\newcommand{\ngood}{$N_{good} ~$}
\newcommand{\nbad}{$N_{bad} ~$}
\newcommand{\nelite}{$N_{elite} ~$}
\newcommand{\nsteps}{$N_{steps} ~$}
\begin{document}

%\runauthor{Oliveira, Silva}

%=========================================
%=========TITLE===========================
%\begin{frontmatter}

\title{A global optimization method for Landau gauge fixing in Lattice QCD}

\author{O. Oliveira\thanks{\textit{email}:orlando@teor.fis.uc.pt},
        P. J. Silva\thanks{\textit{email}:psilva@teor.fis.uc.pt} \\
        Centro de Física Computacional \\ 
        Departamento de Física \\  
        Universidade de Coimbra \\
        3004 - 516 Coimbra, Portugal}

\maketitle

\begin{abstract}
An algorithm for gauge fixing to the Landau gauge in the fundamental modular
region in lattice QCD is 
described. The method, a combination of an evolutionary algorithm with
a steepest descent method, is able to solve the problem of the
nonperturbative gauge fixing. The performance of the combined
algorithm is investigated on $8^4$, $\beta \, = \, 5.7$, 
and $16^4$, $\beta \, = \, 6.0$, lattice SU(3) gauge configurations.
\end{abstract}

%\begin{keyword}
Lattice QCD, Landau Gauge, Gauge Fixing, Gribov Copies

%\end{keyword}

%\end{frontmatter}

%==================================================
\section{Introduction and Motivation}

Quantum Chromodynamics (QCD) is the theory that describes the interaction 
between quarks and gluons. From the dynamical point of view, it is 
usual to separate the high energy regime from the low energy regime. While the
high energy limit of QCD is well described by perturbative methods, 
perturbation theory can not answer a number of important questions. Certainly,
it is not applicable to the low energy limit of QCD. Presently, we do not have
yet an analytical method to tackle this dynamical regime. 
The solution is to solve QCD on the computer \cite{latQCDbook}, where
continuum euclidean space-time is replaced by a discrete set of points, 
the lattice. 
Typical lattices are hypercubes where points are separated by $a$ in each 
direction. 

In the lattice formulation of QCD, the gluon fields $A^a_\mu$ are replaced by 
the links, defined as
\begin{equation}
  U_\mu (x) ~ = ~ \exp \left( i a g_0 A_\mu (x + a \hat{e}_\mu / 2) \right)
  \, ,
\end{equation}
where $\hat{e}_\mu$ are unit vectors along $\mu$ direction. The links
are elements of the SU(3) group. 

QCD is a gauge theory, therefore the fields related by gauge transformations
\begin{equation}
  U_\mu (x) ~ \longrightarrow ~ g(x) ~ U_\mu (x) ~ 
 g^\dagger (x + a \hat{e}_\mu) \, ,
 \hspace{1cm} g \in SU(3) \, ,
\end{equation}
are physically equivalent. The set of fields related by gauge transformations
defines a gauge orbit. From the definition of gauge orbit, it follows that to
study such type of theories it is enough to pick one field from each of the 
orbits.
The identification of one field in each gauge orbit is called gauge fixing.

On the continuum, the problem of the quantisation of gauge theories was solved
long ago by Feynman\cite{feyman}, DeWitt \cite{dewitt} and Faddeev and Popov
\cite{faddeev_popov}. The quantisation method requires a choice of a gauge
condition, uniquely satisfied in each gauge orbit, to define the generating
functional for the Green's functions. For the Landau or the Coulomb gauge and
for small field amplitudes, the gauge condition is uniquely satisfied in each
gauge orbit. However, if large field amplitudes are involved, the gauge fixing
condition has multiple solutions in each gauge orbit \cite{gribov,sciuto}, 
the Gribov
copies, i.e. the nonperturbative quantisation of Yang-Mills theories can not be
described by the usual me\-thods of perturbation theory. This result due to
Gribov for the Coulomb and Landau gauge was generalized by Singer. 
In \cite{singer}, Singer proves that it is impossible to find a local
continuous and unambigously gauge fixing condition for any SU(N) gauge theory
defined on the manifold $S_4$. Singer's theorem was extended to the four-torus
by Killingback \cite{killingback}. 

For the continuum formulation of QCD, in \cite{Zwanziger03} it was argued that
the Landau gauge Faddeev-Popov formula
$\delta ( \partial A) ~ \det [ - \partial \cdot D(A) ] ~ 
\exp [ - S_{YM} (A) ]$, restricted to the region where the Faddeev-Popov 
operator is positive $- \partial \cdot D(A) > 0$ (Gribov region), provides
an exact non-perturbative quantisation for QCD.

The lattice formulation of gauge theories does not require gauge fixing. 
However, gauge fixing is necessary to study the Green's functions of
the fundamental fields like, for example, the gluon and quark propagators and
the quark-gluon vertex. The propagators contain information about
the mechanisms of confinement \cite{Zwanziger91}
and chiral symmetry breaking \cite{Al03a,Al03c}. The quark-gluon
vertex allows a first principles determination of the running coupling constant
of QCD \cite{davies03}. In addition, by choosing a gauge one can compute
renormalisation constants for composite operators by sandwiching the operators
between quark states \cite{martinelli95}. At least for the usual gauges like
the Landau and Coulomb gauges, the lattice studies that rely on a gauge
fixing condition have a fundamental problem: how to properly define a
nonperturbative gauge fixing condition, i.e. how to eliminate the influence
of the different Gribov copies on the results. The implications of
\cite{Zwanziger03} for the lattice formulation of QCD remain to be 
investigated.

In this paper we consider the problem of gauge fixing for the Landau gauge.
On the lattice, Landau gauge fixing can be viewed as a global optimization
problem \cite{giusti}. Typically, we have a minimizing function with many 
local minima, the Gribov copies, and, to eliminate the ambiguities related to
the various minima, we aim to find the absolute minimum. 
In this work, the gauge defined by the absolute minimum of the 
minimizing function is named minimal Landau gauge.

We report on an algorithm that combines a local optimization
method\footnote{By local optimization method we mean an algorithm that seeks 
only a local solution, i.e. a point at which the function is smaller than all 
other points in its vicinity.}, 
a Fourier accelerated steepest descent \cite{davies}, 
with an implementation of an evolutionary algorithm \cite{genalg}, suitable for
global optimization problems\footnote{By global optimization we understand
the problem of computing the absolute minimum/maximum of a given function.}, 
to address the question of gauge fixing in
the minimal Landau gauge. Our investigation shows that a proper combination of
local and global methods identifies the global minimum of the optimizing
function and, in this way, solves the problem of the nonperturbative gauge
fixing. This paper is a full report of the work started in \cite{OlSi01}. 

The paper  is organised as follows. On section 2, the minimal Landau gauge is
defined. In section 3 the local optimization method, the global optimization
method and the combined local+global method are described. 
Section 4 reports on the performance of combined method for $8^4$ and
$16^4$ lattices. Finally, conclusions and discussion are given in section 5.

%==========================================================================
%==========================================================================
\section{The Minimal Landau Gauge}

On the continuum, the Landau gauge is defined by
\begin{equation}
  \partial_\mu A_\mu ~ = ~ 0 \, .\label{landau_cont}
\end{equation}
This condition defines the hyperplane of transverse configurations
\begin{equation}
\Gamma ~ \equiv ~ \{A: ~ \partial \cdot A \, = \, 0 \} ~ .
\end{equation}
It is well known \cite{gribov} that $\Gamma$ includes more than one
configuration from each gauge orbit. In order to try to solve the
problem of the nonperturbative gauge fixing, Gribov suggested the use of
additional conditions, namely the restriction of physical configurational 
space to the region
\begin{equation}
   \Omega  ~ \equiv  ~ \{ A:~ \partial\cdot A \, = \, 0,~ 
                              \textit{M}[A] \, \geq \, 0 \} ~ \subset ~ \Gamma
 \, ,
\end{equation}
where $\textit{M}[A] ~ \equiv ~ - \nabla \cdot D[A] $ is the Faddeev-Popov
operator. However, $\Omega$ is not free of Gribov copies and does not provide 
a proper definition of physical configurations.

A suitable definition of the physical configurational space is given by the
fundamental modular region $\Lambda  \subset  \Omega$, the set of the absolute 
minima of the functional
\begin{equation}
   F_{A}[g] ~ =  ~ \int d^{4}x ~ \sum_{\mu} ~ 
    \mbox{Tr}\left[A_{\mu}^{g}(x)A_{\mu}^{g}(x)\right] \, .
 \label{fcont}
\end{equation}
The fundamental modular region $\Lambda$ is a convex manifold \cite{semyonov}
and each gauge orbit intersects the interior of $\Lambda$ only once
\cite{GAnt91,baal92}, i.e. its interior consists of non-degenerate
absolute minima. On the boundary $\partial \Lambda$ there are degenerate 
absolute minima, i.e. different boundary points are Gribov copies of each 
other \cite{baal92,baal94,baal95}. The interior of $\Lambda$, the region of
absolute minima of (\ref{fcont}),  identifies a region free of Gribov 
copies. To this choice of gauge we call the minimal Landau gauge.

On the lattice, the situation is similar to the continuum theory
\cite{zwanziger92,zwanziger94,cucc9711024}. The interior of $\Lambda$ consists
of non-degenerate absolute minima of the lattice version of (\ref{fcont})
and Gribov copies can occur at the boundary $\partial \Lambda$. For a finite
lattice, the boundary $\partial \Lambda$, where degenerate minima may occur,
has zero measure and the presence of these minima can be ignored 
\cite{zwanziger94}.

On the lattice, the Landau gauge is defined by maximising the functional
\begin{equation}
   F_{U}[g] ~ = ~ C_{F}\sum_{x,\mu} \, \mbox{Re} \, \{ \, \mbox{Tr} \,
       [ g(x)U_{\mu}(x)g^{\dagger }(x+\hat{\mu}) ] \, \}  \label{f}
\end{equation}
where 
\begin{equation}
   C_{F}  ~ = ~ \frac{1}{N_{dim}N_{c}V}
\end{equation}
is a normalization constant, $N_{dim}$ is the dimension of space-time, 
$N_{c}$ is the dimension of the gauge group and $V$ represents the lattice 
volume. Let $U_\mu$ be the configuration that maximises $F[g]$ on a given
gauge orbit. For configurations near $U_\mu$ on its gauge orbit, we have
\begin{eqnarray}
 F_U [ 1 + i \omega (x)]  \approx F_U [ 1 ]  +  
            \frac{C_F}{4} \sum_{x,\mu}  i \omega^a (x) 
            \mbox{Tr} \Big[ & &
                      \lambda^a \left( U_\mu (x) \, - \,
                                         U_\mu ( x - \hat{\mu}) \right)
                            ~ -  \nonumber \\
        & &
                            \lambda^a \left( U^\dagger_\mu (x) \, - \,
                                              U^\dagger_\mu ( x - \hat{\mu}) 
                                    \right)
\Big] \, ,
\end{eqnarray}
where $\lambda^a$ are the Gell-Mann matrices. By definition, $U_\mu$ is a 
stationary point of $F$, therefore 
\begin{eqnarray}
\frac{\partial F}{\partial \omega^a (x)} ~ = ~
\frac{i \, C_F}{4} \sum_{\mu} 
            \mbox{Tr} \Big[ & &
                      \lambda^a \left( U_\mu (x) \, - \,
                                         U_\mu ( x - \hat{\mu}) \right)
                            ~ -  \nonumber \\
        & &
                            \lambda^a \left( U^\dagger_\mu (x) \, - \,
                                              U^\dagger_\mu ( x - \hat{\mu}) 
                                    \right)
\Big] ~ = ~ 0 \, .
 \label{statf}
\end{eqnarray}
In terms of the gluon field, this condition reads
\begin{equation}
  \sum_{\mu} \mbox{Tr} \Big[ ~ \lambda^a 
                               \left( A_{\mu}(x + a \hat{\mu}/2) - 
                                      A_{\mu}(x - a \hat{\mu}/2)
                               \right) ~\Big] \, + \, 
   \mathcal{O} (a^2) ~ =  ~ 0 \, ,
\end{equation}
or
\begin{equation}
  \sum_{\mu} \partial_\mu A^a_{\mu}(x) ~ + ~
   \mathcal{O} (a) ~ =  ~ 0 \, ,
\end{equation}
i.e. (\ref{statf}) is the lattice equivalent of the continuum Landau gauge 
condition. The lattice Faddeev-Popov operator $M(U)$ 
is given by the second derivative of (\ref{f}).

Similarly to the continuum theory, on the lattice one defines the region of
stationary points  of (\ref{f})
\begin{equation}
   \Gamma ~ \equiv ~ \{U: ~ \partial \cdot A(U) \, = \, 0 \} \, ,
\end{equation}
the Gribov's region $\Omega$ of the maxima of (\ref{f}),
\begin{equation}
   \Omega ~ \equiv ~ \{U: ~ \partial \cdot A(U)=0 ~ \mbox{and} ~ M(U) \geq 0 \}
\end{equation}
and the fundamental modular region $\Lambda$ defined as the set of the 
absolute maxima of (\ref{f}). The lattice minimal Landau gauge chooses from
each gauge orbit, the configuration belonging to the interior of
$\Lambda$.

The evidence for lattice Gribov copies, i.e. different maxima of $F_U$,
was established long time ago  \cite{nakamura,forcrand91,marinari} but their 
influence on physical observables is not clear. For the lattice Landau gauge,
SU(2) simulations suggest that the influence of Gribov copies is at the level
of the simulation statistical error \cite{cucchieri97,FuNa03}. 
For SU(3) there is no
systematic study but it is believed that the Gribov noise is contained within
the statistical error of the Monte Carlo. Here, we will not discuss the role
of Gribov noise on correlator functions but an algorithm
for finding the absolute maximum of $F_U [g]$. For a discussion on the
influence of Gribov copies on the gluon propagator see 
\cite{cucchieri97,FisHad03,nosso}.

%=========================================================================
%=========================================================================
\section{Optimization Methods}

The algorithm for minimal Landau gauge fixing reported in this paper combines
a local and a global optimization method. For completeness, in this section we
outline the local method and describe the global and combined 
local+global algorithms.

On the gauge fixing process, the quality of the gauge fixing is measured
by
\begin{equation}
  \theta ~ = ~ \frac{1}{VN_{c}} ~ \sum_{x} \mbox{Tr}
      [\Delta(x)\Delta^{\dag}(x)] \label{theta}
\end{equation}
where 
\begin{equation}
 \Delta(x) ~ = ~ \sum_{\nu} \left[ U_\nu ( x - a \hat{e}_\nu) \, - \,
                                   U^\dagger_\nu (x) \, - \, \mbox{h.c.}
                                   \, - \, \mbox{trace} \right]
\end{equation}
is the lattice version of $\partial_\mu A_\mu \, = \, 0$.

%=========================================================================
%=========================================================================
\subsection{Local Optimization}

By definition, a local optimization method computes a local maximum of 
$F_U [g]$. For Landau gauge fixing, a popular local optimization method is
the steepest descent \cite{davies} method.

The naive steepest descent method faces the problem of critical slowing down
when applied to large lattices. Critical slowing down can be reduced by
Fourier acceleration. In the Fourier accelerated method, in each iteration 
one chooses
\begin{equation}
   g(x) ~ =  ~ \exp \Bigg[ \hat{F}^{-1} ~ 
                               \frac{\alpha}{2} \,
                               \frac{ p_{max}^{2} a^{2}}{ p^{2} a^{2} } ~
                               \hat{F} ~ 
           \left( \sum_\nu  \Delta_{- \nu} \left[ U_\nu ( x ) \, - \,
                                          U^\dagger_\nu ( x ) \right]
                            \, - \, \mbox{trace} \right) \Bigg]
\label{sd}
\end{equation}
where
\begin{equation}
\Delta_{-\nu} \left( U_\mu (x) \right) ~ =  ~ 
       U_\mu ( x - a \hat{e}_\nu) \, - \, U_\mu ( x ) \, ,
\label{delta}
\end{equation}
$p^{2}$ are the eigenvalues of $(-\partial^{2})$, $a$ is the lattice spacing
and $\hat{F}$ represents a fast Fourier transform (FFT). For the parameter
$\alpha$ we use the value 0.08 \cite{davies}. For numerical purposes, it is
enough to expand to first order the exponential in (\ref{sd}), followed by a 
reunitarization of $g(x)$.

For large lattices (\ref{sd}) is not the best way to solve the problem of
critical slowing down. In \cite{CuMe98a,CuMe98b} 
a method was developed that avoids
the use of FFT, has a dynamical critical exponent close to zero and 
the advantage to be easily parallelized. 
In this work we use the Fourier accelerated steepest descent method (SD).

%======================================================================
%======================================================================
\subsection{Global Optimization}

Global optimization methods aim to find the absolute maximum or minimum of a
multidimensional function. Presently, there is not a method that can assure, 
with certainty, that the computed maximum in a single run is the absolute 
maximum. Simulated annealing (SA) is, probably, the most popular method for
global optimization. However, evolutionary algorithms (EA) \cite{genalg}
are an alternative to simulated annealing. 
The ``advantage'' of evolutionary algorithms relatively to SA is that EA work
with multiple candidates for maximum/minimum
in a single run and, in principle, 
can avoid or reduce the number of multiple runs necessary to identify the 
global optimum. For us, this provided the motivation to try the use of EA for
gauge fixing in lattice QCD.

Evolutionary algorithms (EA) are a generalization of genetic algorithms (GA).
Genetic algorithms are inspired in natural selection and in the theory of
evolution of species.
The language spoken in evolutionary programming is borrowed from genetics.
The vector of the parameters to optimize is called chromosome or individual.
A population consists in a number of individuals. The function to optimize
is the cost function.

For the gauge fixing problem, a chromosome is the set of matrices $g(x)$
that defined a gauge transformation. The cost function is the functional
$F_U [ g ]$.

An evolutionary code starts generating a set of tentatives of solutions,
the initial population. In the following the number of individuals in the
initial population will be refered by \nipop$\!\!$. In our case, the initial 
population was generated randomly. After sorting the initial population 
according to their cost function value, \npop members were selected, using
a roullette-wheel method \cite{genalg},
to begin the evolutionary phase. 
The number of individuals in the population
was always kept fixed to \npop$\!\!$. In this work we used
\nipop/\npop = 2.5.

The population evolution was performed according to the rules
\begin{enumerate}
\item the best \ngood individuals survive for the next generation;

\item \nbad = \npop - \ngood are replaced by new chromosomes. The new
individuals are generated by reproducing the \ngood members of the population.
In this work we set \ngood = \npop/2.

\item For mating or reproduction two good chromosomes are selected and
give ``birth'' to two offsprings. The process completes after generation of
\nbad new offsprings. In this work parents were selected using the so-called
roulette-wheel selection \cite{genalg}, a method which favours the best 
chromosomes in the population.

\item A new generation is defined only after mutating the population 
cons\-tructed after point 3. No mutation was applied to the best member of
the population.
\end{enumerate}

In order to reproduce, a population requires a set of rules to make childs
from the parents, the genetic operators. In this work we considered
the following operators
\begin{itemize}
\item Random Crossover (RC)

For evolutionary and genetic algorithms, crossover is a fundamental mating 
operator that mimics the crossover observed in biological systems: after 
selection of a set of contiguous genes in the chromosomes, the two childs are
built by interchanging the chosen piece of the parents genetic material. Our 
implementation of the crossover is slightly different. On the lattice we
select randomly $V \times p_{intcross}$ points; the random variable 
$p_{intcross}$ takes values in $[0.40, \, 0.70]$. 
The offsprings are defined by interchanging the
parents matrices $g$ at these points\footnote{In literature this type of
crossover is also known as uniform crossover.}. Note that crossover does not
implies creation of new genes.

\item Random Blending (RB)

Blending operators try to overcome the crossover problem of gene creation.
Our implementation of blending starts by choosing a set of lattice points
similarly as in RC. For the first child, we select a random value\footnote{The
choice $\beta \in [0,1]$ is the most simple blending method, but has the 
disadvantage that it does not create new values outside the interval defined
by the parents genes.} for $\beta \in [0,2]$. 
For the selected points, the $g$ matrices are given, after reunitarization, by
\begin{equation}
 \beta \, g_{1} ~ + ~ (1 \, - \,\beta) \, g_{2} \, ,
\end{equation}
where $g_1$ and $g_2$ denote parents. In the remaining
lattice points we set $g ~ = ~ g_1$. The genetic material of the second
offspring is generated in the same way. The difference being that $\beta$
is choosen different at each of the selected lattice points and in the 
remaining points we set $g ~ = ~ g_2$.
\end{itemize}
Each mating operator has an associated probability. After parent selection,
it is tested if the chosen parent is able to reproduce by comparing an
uniformly distributed random number in $[0, \, 1]$ with the mating probability.

The mating operators just recombine the genetic information of the population.
To explore more effectively the cost surface, an evolutionary code apply
mutation operators after the mating phase. 
These operators change a few genes of a chromosome either by replacing the
gene by a neighbouring value or replacing the gene by a completely 
different value. In this work, we considered the following mutation operators
\begin{itemize}
\item Addition mutations (MA)
\begin{equation}
 g(x)  ~ \longrightarrow ~ g(x) \, + \, \epsilon A \, ,
\end{equation}

\item Substitution mutations (MS)
\begin{equation}
  g(x) ~  \longrightarrow ~  A \, ,
\end{equation}

\item Expansion mutations (ME)
\begin{equation}
 g(x)  ~ \longrightarrow  ~  g(x) \, \left( 1 \, + \, \epsilon A \right)
\end{equation}
\end{itemize}
where $\epsilon$ ($\left| \epsilon \right| \le 0.025$) is a random number and
$A$ is a random SU(N) matrix. The resulting matrix is properly reunitarized.
Each mutation operator is applied to all population skipping the best
\nelite individuals (in our work \nelite = 1). Like for the mating operators, 
mutation have an associated probability too. For each operator, and for
each individual, we go through the lattice and apply the operator in the
corresponding $g$ matrices according to the respective probability.

Each complete iteration of the algorithm (selection, mating and mutation) is
called generation. 

The probabilities associated to each genetic operator were defined to 
ma\-xi\-mi\-ze
the performance of the pure evolutionary code. The large number of parameters
makes a detailed study of the probabilities quite hard to perform. However,
for practical purposes, we used the procedure described below to define our
algorithm. Set all probabilities to zero except for MS. For MS take 0.01 for
the probability. Change the probability of RC and choose the value that 
optimizes the performance of the algorithm. After setting the probability for
RC, repeat the procedure for MA, then for RB, then for ME. Finally, check for
the value of the MS probability.  The probabilities associated to the genetic 
and mutation operators used in our study are
\begin{center}
\begin{tabular}{||c@{\hspace{0.5cm}}r||}
\hline\hline
    Random Crossover  &  $p_{rc} ~ = ~ 0.40$  \\
    Random Blending   &  $p_{rb} ~ = ~ 0.70$  \\
    Addition Mutation &  $p_{ma} ~ = ~ 0.04$ \\
    Substitution Mutation & $p_{ms} ~ = ~ 0.04$ \\
    Expansion Mutation    & $p_{me} ~ = ~ 0.02$ \\
\hline\hline
\end{tabular}
\end{center}

In what concerns Landau gauge fixing, the performance of the pure
evolutionary algorithm for Landau gauge fixing was quite disapointing; the
best run ended with $\theta  \, \sim \, 10^{-1}$. 
This can be understood as a consequence of the large dimension of the problem
and of the nature of $F_U$. In conclusion, the performance of the pure 
evolutionary algorithm for the 4D Landau gauge fixing problem is similar to 
the performance observed in simplified versions of the problem 
\cite{yamaguchi,australianos}.

%==========================================================================
%==========================================================================
\subsection{Combined Global + Local Optimization}

For minimal Landau gauge fixing in lattice QCD, the global optimization 
pro\-blem can be overcomed by combining the local and the global algorithms
described above. From the point of view of the evolutionary algorithm, a
possible combined algorithm means redefining the cost function as
\begin{equation}
   f[g; N] ~ = ~ F_U[g] \mbox{ after } N \mbox{ local steepest descent steps}.
 \label{costfunction}
\end{equation}
As described below, with a proper choice of $N$ it is possible to identify the
global maximum of $F_U$.

%=====================================================================
\section{Results for Combined Algorithm and the Minimal Landau Gauge}

The combined algorithm was studied with SU(3) gauge configurations on
$8^4$ ($\beta \, = \, 5.7$) and $16^4$ ($\beta \, = \, 6.0$) lattices. The
gauge configurations were generated with the MILC code \cite{milc} using a
combination of four over-relaxed and one Cabibbo-Marinari updates, with a 
separation between configurations of 3000 combined updates. For each of the 
lattices, the combined algorithm was investigated in detail for at least three
configurations.

For Landau gauge fixing, the absolute maximum of $F_U$ was computed by running,
for each gauge configuration, 1000 local algorithms for the smaller lattice 
and 500 on the larger lattice, starting from different random chosen points.
A local minimum was defined by demanding that $\theta \, < \, 10^{-10}$ for
the smaller lattice and $\theta \, < \, 10^{-15}$ for the larger lattice.
The candidate for absolute maximum computed with the combined algorithm was
compared with the candidate for absolute maximum from the multiple local
algorithm runs. In all the simulations, we never observed a larger maximum than
the one obtained with the multiple runs of the steepest descent method.
Preliminary results on  the performance of the combined algorithm were given 
in \cite{OlSi01}.

%=========================================================================
\subsection{$8^4$ lattices}

For the $8^4$ lattice, 10 gauge configurations were generated. 
The study of their Gribov copies structure was performed by running 1000 SD
on each of the configurations. Then, a detailed study of the three
configurations with the largest number of maxima was performed as described
below.

\begin{figure}[t]
\begin{center}
\includegraphics[angle=-90,scale=0.75]{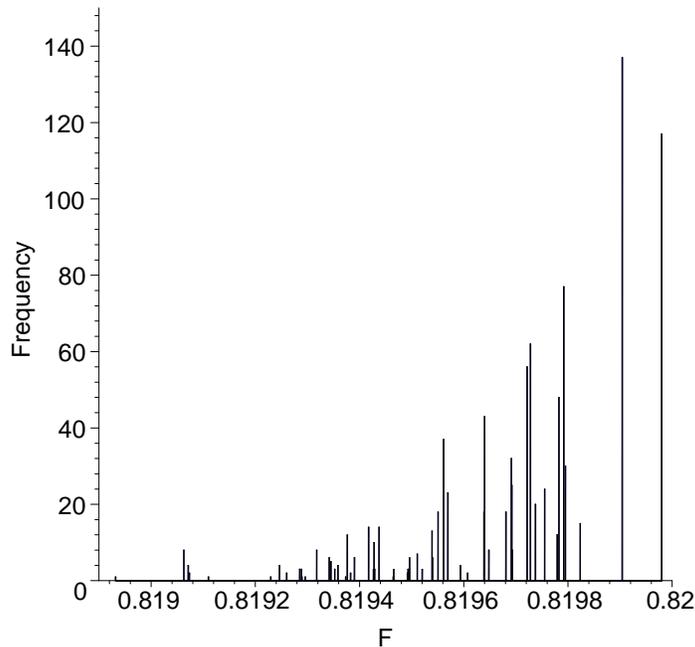}
\end{center}
\caption{Local maxima of one of the $8^{4}$ SU(3), $\beta=5.7$, configurations
after 1000 \textit{Steepest Descent} starting from random points 
($\theta \leq 10^{-10}$).}
\label{su3_fmax_milc1}
\end{figure}

The number of local maxima computed in the multiple runs with the steepest
descent method was quite large. Figure \ref{su3_fmax_milc1} resumes the 
1000 SD runs for one of the configurations used to test the algorithm. 
The figure shows not only a large number of local maxima, the Gribov copies, 
but also that the most probable maxima are associated with the largest values
of $F_U$. Note that, for the configuration considered, the copy with the 
largest frequency is not the absolute maximum. These properties are a general 
trend observed for some of the configurations.

%===========================================================================
%===========================================================================
%===========================================================================
%===========================================================================

The combined evolutionary algorithm steepest descent method (CEASD) has a 
large number of parameters and to
establish the algorithm we tried to cover, as much as possible, the space of
parameters. Table \ref{pop_val} is a summary of the various runs. 
All results reported in this paper, for this smaller lattice, consider 
runs with 400 generations and use 
\nipop = $2.5 ~ \times ~$\npop, \ngood = \npop/2 and \nelite = 1.

%===========================================================================
\begin{table}[t]
\begin{center}
\begin{tabular}{||l|c|c|c|c|c||}
\hline\hline
\nipop & \textbf{10}  & \textbf{20} & \textbf{30} & 
                   \textbf{40} & \textbf{50} \\
\hline
\npop &    4&8&12&16&20 \\
\hline
\ngood &    2&4&6&8&10 \\
\hline\hline
\end{tabular}
\caption{Evolutionary populations considered on the $8^4$ study. The number
of generations used in each run was 400.} \label{pop_val}
\end{center}
\end{table}
%============================================================================

\begin{figure}[t]
\begin{center}
\includegraphics[angle=0,origin=c,scale=0.7,bb=58 251 536 591]
{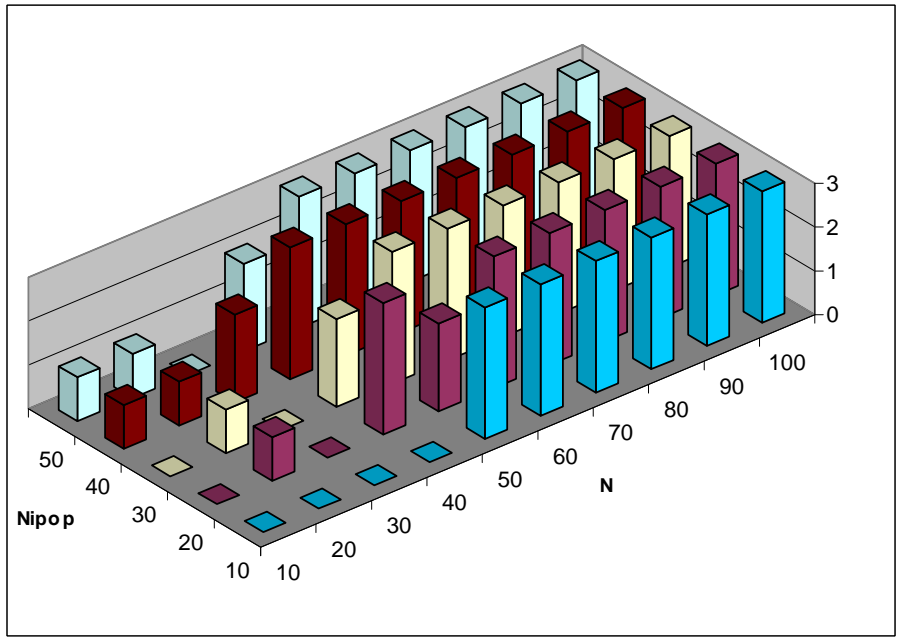}
\caption{Number of successful runs for the combined method, for the three 
$8^4$ test configurations, as function of \nipop and $N$.}
\label{estmilc_histogram}
\end{center}
\end{figure}

For the combined algorithm, we observed that by increasing $N$ in 
(\ref{costfunction}), the computed maximum, i.e. the maximum computed after
applying a SD to the best member of population of the last generation, becomes 
closer to the absolute maximum. 
Moreover, there is a minimum number of $N$, \nsteps, 
such that the computed maximum of CEASD is the absolute maximum of $F_U$. 
Figure \ref{estmilc_histogram} reports the number of successful runs of the
combined
method, for the three $8^4$ test configurations, as function of \nipop and $N$.

\begin{figure}[t]
\begin{center}
\includegraphics[angle=-90,scale=0.8]{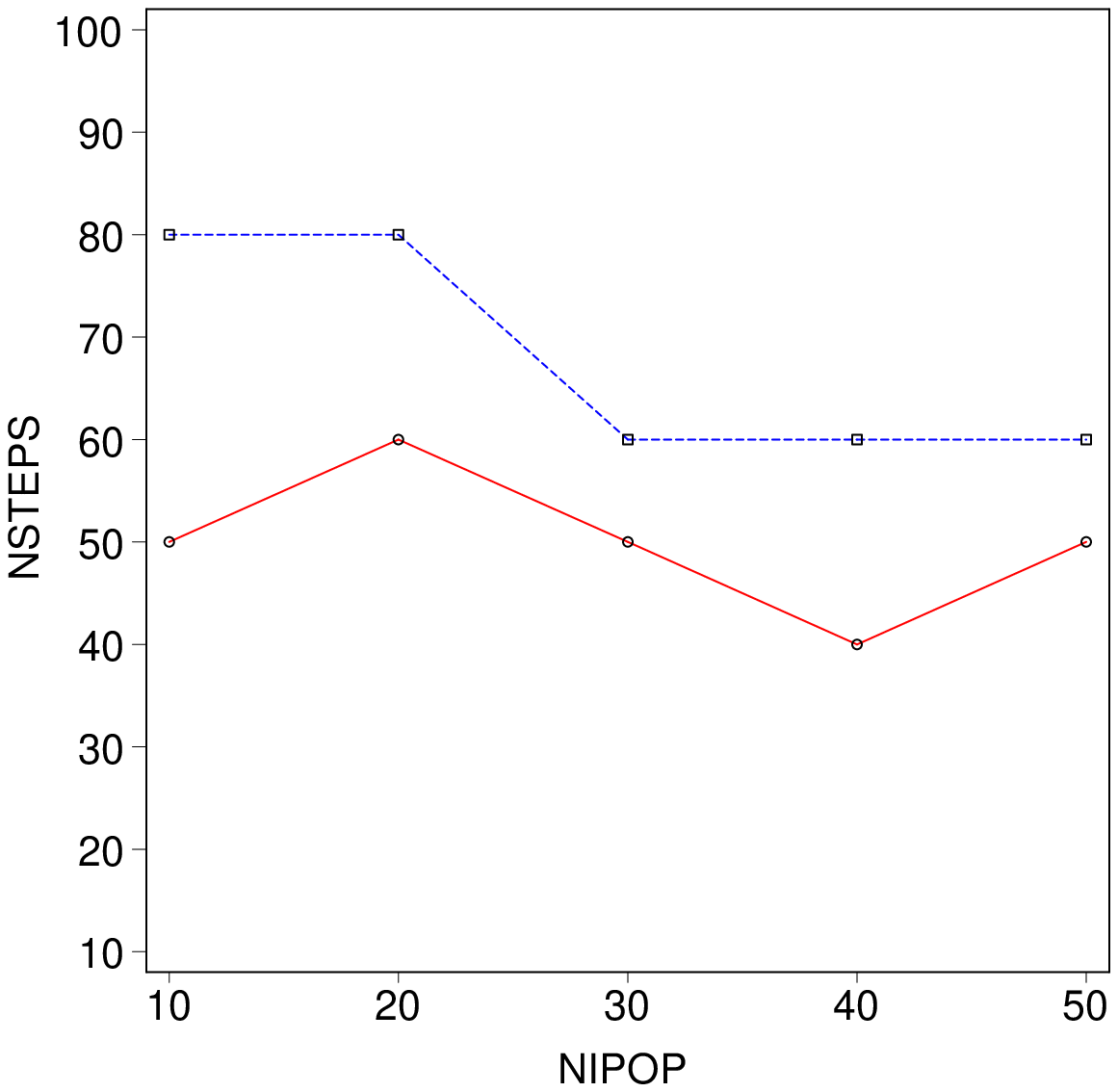}
\end{center}
\caption{\nsteps versus \nipop for $8^{4}$ SU(3), $\beta=5.7$,
configurations. The solid lines gives \nsteps for 400 generations. The
dashed line is \nsteps for 50 generations.}\label{nvp_milcv2}
\end{figure}

Figure
\ref{estmilc_histogram} shows that, for each population size, there is a
minimum value of $N$, \nsteps$\!\!$, such that the CEASD algorithm correctly 
computes the absolute maximum. 
Figure \ref{nvp_milcv2} shows \nsteps as a function of the initial population.
The solid line is \nsteps for 400 generations and the dashed line is the value
of $N$ required to identify the absolute maximum in 50 generations. Results
seem to suggest that for larger populations, \nsteps should become
smaller.

%=========================================================================
%=========================================================================
%========================================================================
%========================================================================
%========================================================================
%========================================================================

Table \ref{estmilcv2_generations} reports the first generation that includes 
the absolute maximum in the population\footnote{We monitored the presence
of the absolute maximum each 50 generations.}. The results seems
to suggest that, for an $8^4$ lattice, 50 generations may be a safe number
of generations for the CEASD algorithm to compute the absolute
maximum of $F_U$.

\begin{center}
\begin{figure}[t]
  \subfigure[\nipop=10]{ \label{ev_milc8_1:n10}
  \begin{minipage}[b]{0.48\textwidth}
    \centering
    \includegraphics[angle=-90,origin=c,scale=0.6]{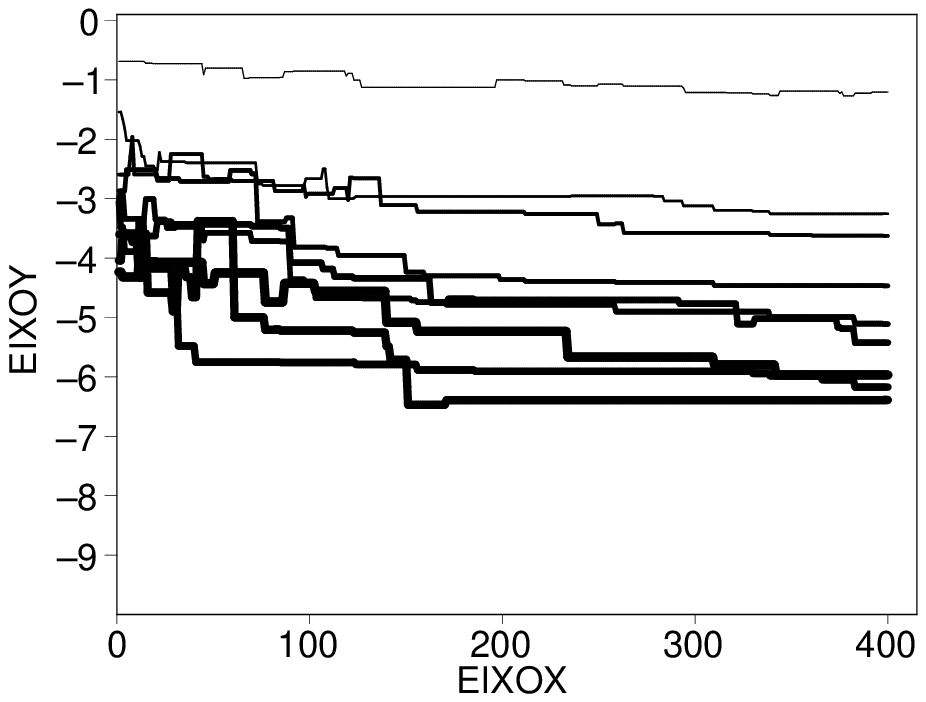}
  \end{minipage} }
  \subfigure[\nipop=20]{ \label{ev_milc8_2:n20}
  \begin{minipage}[b]{0.48\textwidth}
    \centering
    \includegraphics[angle=-90,origin=c,scale=0.6]{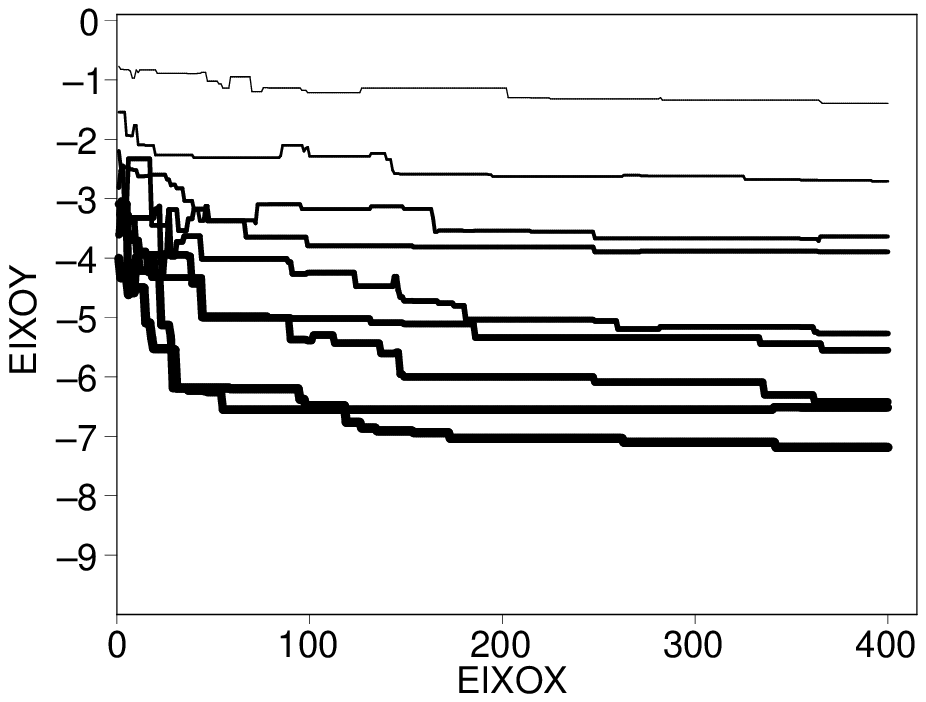}
  \end{minipage} }\\[20pt]

  \subfigure[\nipop=40]{ \label{ev_milc8_3:n40}
  \begin{minipage}[b]{0.48\textwidth}
    \centering
    \includegraphics[angle=-90,origin=c,scale=0.6]{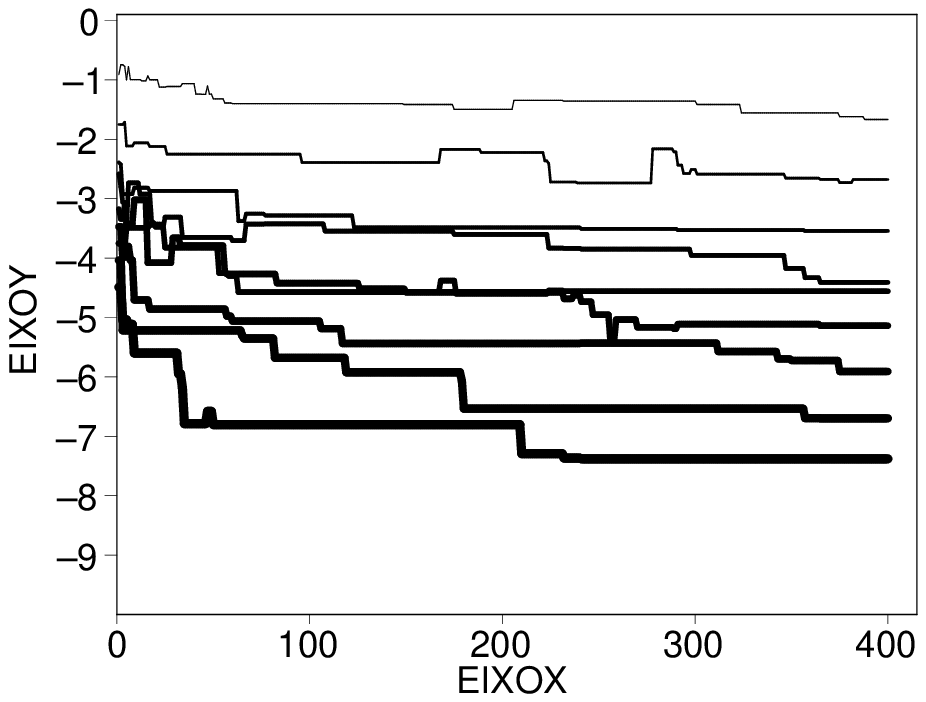}
  \end{minipage} }
  \subfigure[\nipop=50]{ \label{ev_milc8_4:n50}
  \begin{minipage}[b]{0.48\textwidth}
    \centering
    \includegraphics[angle=-90,origin=c,scale=0.6]{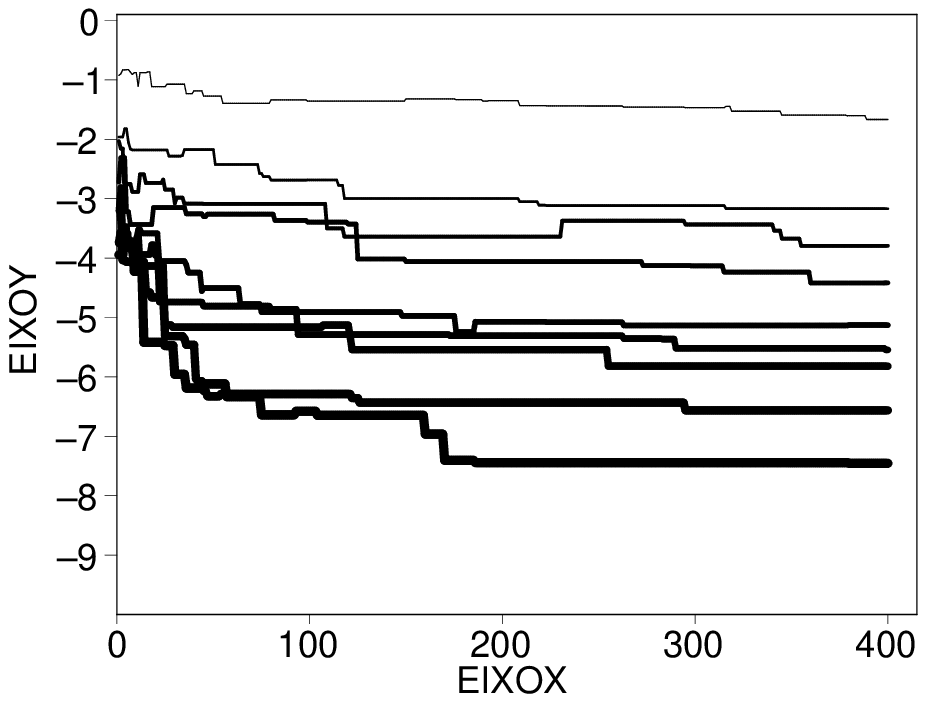}
  \end{minipage} }\\[20pt]
  \caption{$\log(\theta)$ for one of the $8^4$ test configurations and 
different population sizes. The value of N is represented by the thickness 
of the line (larger thickness meaning larger $N$).} \label{ev_milc8_1}
\end{figure}
\end{center}

Figure \ref{ev_milc8_1} reports, for different population sizes, typical
evolutions of $\theta$ for one of the tested configurations. They show that,
in each run, $\theta$ decreases rapidly in the first generations, with
its value decreasing by roughly 3 to 4 orders of magnitude in the first 50
generations, and then remaining approximately constant.
Moreover, in order to properly identify the absolute
maximum of $F_U$, the algorithm seems to require 
$\theta ~ \sim ~ 10^{-6} - 10^{-7}$ after generation 50.

%========================================================================
%========================================================================
\psfrag{EIXOX}{Generation}
\psfrag{EIXOY}{$\log(\theta)$}
%========================================================================
%========================================================================

%======================================================
%======================================================
\begin{table}[t]
\begin{center}
\begin{tabular}{||c||c|c|c|c|c||}
\hline\hline
   & \multicolumn{5}{|c||}{ \nipop} \\
\hline
   \nsteps & \textbf{10}  & \textbf{20} & \textbf{30} & 
    \textbf{40} & \textbf{50} \\
\hline
     \textbf{10} &    -&-&-&-&- \\
     \textbf{20} &    -&-&-&-&- \\
     \textbf{30} &    -&-&-&-&- \\
     \textbf{40} &    -&100&-&100&-\\
     \textbf{50} &    150&-&150&150&150\\
     \textbf{60} &    150&50&50&50&50 \\
     \textbf{70} &    350&100&50&50&50 \\
     \textbf{80} &    50&50&50&50&50 \\
     \textbf{90} &    50&50&50&50&50 \\
    \textbf{100} &    50&50&50&50&50 \\
\hline\hline
\end{tabular}
\caption{Number of generations required by the CEASD algorithm to identify the
absolute maximum for an $8^4$ lattice.}\label{estmilcv2_generations}
\end{center}
\end{table}
%======================================================
%======================================================

In conclusion, for an $8^4$ lattice it is possible to define a set of
parameters such that the CEASD algorithm identifies the
gauge transformation that maximizes $F_U$. For this smaller lattice, our 
choice being \nsteps = 100, \nipop = 10 for runs with 200 generations. 
Note that from figure \ref{nvp_milcv2} one reads \nsteps = 50. However, since
evolutionary algorithms are statistical algorithms and the
combined algorithm requires a relatively low value for $\theta$ to
access the absolute maximum of $F_U$, our choice for \nsteps and
the number of generations was conservative. Indeed, results show that 
similar results can be obtained for runs with only 50 
generations\footnote{We tested running the code on 10 configurations for
\nsteps = 80, \nipop = 10 and for 200 generations. 
Of all the configurations, only one didn't arrive to the
absolute maximum. For \nsteps = 100, of the 10 configurations tested nine
got the absolute maximum in 50 generations and only one required 100 generations
to compute correctly the maximum.}.
Decreasing the number of generations implies either increasing \nsteps
(increasing the computational cost of the cost function),
increasing \nipop (increasing the memory requirements) or 
relying on multiple runs of the algorithm. Of course, the user should choose
between the different possible solutions depending on the computational power
he has available.

In order to get an idea on the cpu time required by the CEASD, we benchmarked
the code on a Pentium IV at 2.40 GHz. For the $8^4$ lattice,
\nsteps = 100, \nipop = 10 and requiring $\theta \, < \, 10^{-15}$ for the
final steepest descent applied to the best member of the population of the
last generation, we measured
\begin{center}
\begin{tabular}{ccc}
     number of generations &  \mbox{   }  & cpu time (s) \\
     50   &  & 2633 \\
     200  &  & 10859
\end{tabular}
\end{center}
meaning that the CEASD algorithm requires about 54 seconds/generation. For the
same gauge fixing precision, the
steepest descent method requires 56 seconds. Therefore, the
time required by a run with 
200 generations is similar to the time
required by 200 multiple steepest descent. 
At this point, a warning should be given to the reader. The CEASD code has 
space for optimization, therefore the cpu times 
reported above should be read as order of magnitudes. The CEASD memory 
requirements for the evolutive phase (\npop = 4) are about 15MB.

In the next section we report on the CEASD algorithm for a larger lattice.

%==========================================================================
%==========================================================================
\subsection{$16^4$ lattices}

%=====================================================================
%=====================================================================
\begin{figure}[t]
\begin{center}
\includegraphics[angle=-90,scale=0.7]{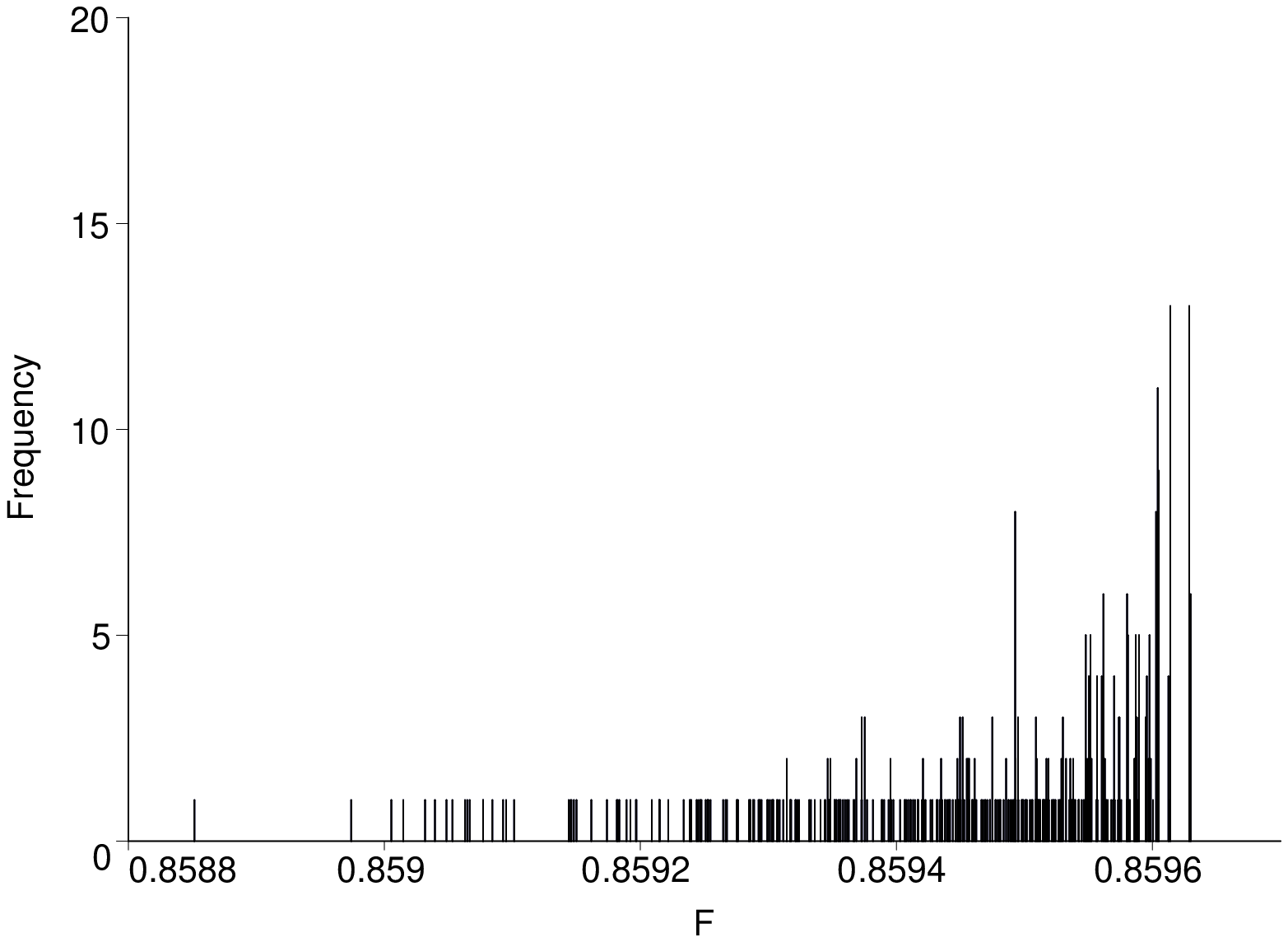}
\end{center}
\caption{Local maxima of a $16^{4}$ SU(3), $\beta=6.0$
configuration achieved after 500 Steepest Descent starting from random 
starting points ($\theta \leq 10^{-15}$). For the three configurations used
to set the algorithm, the number of different Gribov copies found
after 500 SD was 177/500, 238/500, 326/500 for configurations number 66000,
72000 and 9000, respectively.}
\label{su3_fmax_16_9000}
\end{figure}
%=====================================================================
%=====================================================================

For the larger lattice considered in this work, seven $\beta \, = \, 6.0$ 
gauge configurations were generated. Similarly to what was done for the 
$8^4$ lattice, the Gribov copies structure was studied applying 500 steepest 
descents started from different randomly chosen points. In order to test
the algorithm we performed a detailed study for the three configurations with
the largest number of Gribov copies.

The first observation being that the number of local maxima is now much larger
than in the $8^4$ lattice. 
Figure \ref{su3_fmax_16_9000} shows the Gribov copies found in 
500 multiple SD of one of the configurations used in the detailed study of 
the CEASD algorithm. 
A similar figure for a $8^4$ configuration is figure \ref{su3_fmax_milc1}. 
Not only, the number of local maxima increases but also the maxima become 
closer to each other when compared to the smaller lattice. 
To give an idea of the local maxima for the $16^4$ configurations, in table 
\ref{topten16a4} we list the first five highest values of $F_U$ computed 
after the 500 SD method, the 10$^{\mbox{th}}$ and the smaller $F_U$.
From the numerical point of view, this difference makes the global 
optimization problem a much harder problem to solve. Concerning the frequency
of the local maxima, the results for the $16^4$ and $8^4$ lattices are 
similar. The most probable maxima are associated with the largest values of
$F_U$ but the copy with the largest frequency is not always the absolute 
maximum of $F_U$.
%========================================================================
%========================================================================
\begin{table}[t]
\begin{center}
\begin{tabular}{|r||c|r||c|r||c|r|}
\hline
      & Conf66000  & Freq. & Conf72000 & Freq. & Conf9000 & Freq. \\
\hline
     1  & 0.86013650 & 17 &   0.85964596 & 15 &   0.85962982 &  6 \\
     2  & 0.86013552 &  3 &   0.85963938 &  2 &   0.85962880 & 13 \\
     3  & 0.86013533 &  8 &   0.85963928 &  7 &   0.85961392 & 13 \\
     4  & 0.86013430 & 10 &   0.85963888 &  5 &   0.85961286 &  1 \\
     5  & 0.86013269 &  6 &   0.85963756 &  1 &   0.85961242 &  4 \\
    10  & 0.86012155 &  6 &   0.85963328 &  4 &   0.85960036 &  1 \\
smaller & 0.85907152 &  1 &   0.85866489 &  1 &   0.85885161 &  1 \\
\hline
\end{tabular}
\caption{$F_U$ values after 500 SD - $16^4$ lattice.}
\label{topten16a4}
\end{center}
\end{table}
%========================================================================
%========================================================================

For the larger lattice, the investigation of the algorithm did not cover the
same set of parameters as in the study of the smaller lattice. Indeed, due to
the difference on the size of a gauge configuration, a factor of $2^4$, we
only considered the smallest population sizes, namely \nipop = 10, 20. The
larger populations were avoided because of the large memory requirements.
In what concerns the number of generations on each run, for the larger lattice
we only considered runs up to 200 generations and checked for the presence of
to the best maximum each 50 
generations. As in the previous section, in all runs we used 
\nipop = 2.5\npop$\!\!$.

%==========================================================================
%==========================================================================
\begin{table}[t]
\begin{center}
\tiny{
\begin{tabular}{||c|r|c|c|c|c|c|c||}
\hline\hline
  & \nipop & \multicolumn{3}{|c|}{\textbf{10}}   & 
             \multicolumn{3}{|c|}{ \textbf{20}} \\
\hline
$N$ & \textbf{Generation} & 66000  & 72000 & 9000 &   
                            66000   &   72000   & 9000 \\
\hline
\textbf{120} &     50     &  \textbf{1}  &  \textbf{4}  &  \textbf{2}  &     
                             \textbf{1}  &  \textbf{4}  &  \textbf{1}     \\
             &    100     &  \textbf{1}  &  \textbf{3}  &  \textbf{2}  &     
                             \textbf{1}  &  \textbf{1}  &  \textbf{1}     \\
             &    150     &  \textbf{1}  &  \textbf{3}  &  \textbf{2}  &  
                             \textbf{1}  &  \textbf{1}  &  \textbf{1}     \\
             &    200     &  \textbf{1}  &  \textbf{3}  &  \textbf{2}  &  
                             \textbf{1}  &  \textbf{1}  &  \textbf{1}     \\
\hline
\textbf{130} &     50     &  \textbf{1}  &  \textbf{3}  &  \textbf{6}  &
                             \textbf{1}  &  \textbf{2}  &  \textbf{1}     \\
             &    100     &  \textbf{1}  &  \textbf{1}  &  \textbf{2}  &  
                             \textbf{1}  &  \textbf{2}  &  \textbf{1}     \\
             &    150     &  \textbf{1}  &  \textbf{1}  &  \textbf{2}  &  
                             \textbf{1}  &  \textbf{2}  &  \textbf{1}     \\
             &    200     &  \textbf{1}  &  \textbf{1}  &  \textbf{2}  &
                             \textbf{1}  &  \textbf{1}  &  \textbf{1}     \\
\hline
\textbf{140} &     50     &  \textbf{3}  &  \textbf{9}  &  \textbf{2}  &
                             \textbf{1}  &  \textbf{1}  &  \textbf{2}  \\
             &    100     &  \textbf{1}  &  \textbf{9}  &  \textbf{1} &
                             \textbf{1}  &  \textbf{1}  &  \textbf{2}  \\
             &    150     &  \textbf{1}  &  \textbf{6}  &  \textbf{1} &
                             \textbf{1}  &  \textbf{1}  &  \textbf{2}  \\
             &    200     &  \textbf{1}  &  \textbf{1}  &  \textbf{1} &
                             \textbf{1}  &  \textbf{1}  &  \textbf{2}  \\
\hline
\textbf{150} &     50     &  \textbf{3}  &  \textbf{3}  &  \textbf{1} &
                             \textbf{1}  &  \textbf{1}  &  \textbf{1}  \\
             &    100     &  \textbf{1}  &  \textbf{3}  &  \textbf{1} &
                             \textbf{1}  &  \textbf{1}  &  \textbf{1}  \\
             &    150     &  \textbf{1}  &  \textbf{3}  &  \textbf{1} &
                             \textbf{1}  &  \textbf{1}  &  \textbf{1}  \\
             &    200     &  \textbf{1}  &  \textbf{3}  &  \textbf{1} &
                             \textbf{1}  &  \textbf{1}  &  \textbf{1}  \\
\hline
\textbf{160} &     50     &  \textbf{1}  &  \textbf{1}  &  \textbf{2}  &
                             \textbf{2}  &  \textbf{1}  &  \textbf{1}  \\
             &    100     &  \textbf{1}  &  \textbf{1}  &  \textbf{2}  &
                             \textbf{2}  &  \textbf{1}  &  \textbf{1}  \\
             &    150     &  \textbf{1}  &  \textbf{1}  &  \textbf{2}  &
                             \textbf{2}  &  \textbf{1}  &  \textbf{2}  \\
             &    200     &  \textbf{1}  &  \textbf{1}  &  \textbf{2}  &
                             \textbf{1}  &  \textbf{1}  &  \textbf{2}  \\
\hline
\textbf{170} &     50     &  \textbf{1}  &  \textbf{1}  &  \textbf{2} &
                             \textbf{1}  &  \textbf{1}  &  \textbf{1}   \\
             &    100     &  \textbf{1}  &  \textbf{1}  &  \textbf{2} &
                             \textbf{1}  &  \textbf{1}  &  \textbf{1}    \\
             &    150     &  \textbf{1}  &  \textbf{1}  &  \textbf{2} &
                             \textbf{1}  &  \textbf{1}  &  \textbf{1}    \\
             &    200     &  \textbf{1}  &  \textbf{1}  &  \textbf{2} &
                             \textbf{1}  &  \textbf{1}  &  \textbf{1}     \\
\hline
\textbf{180} &     50     &  \textbf{1}  &  \textbf{1}  &  \textbf{1}  &  
                             \textbf{2}  &  \textbf{1}  &  \textbf{2}  \\
             &    100     &  \textbf{1}  &  \textbf{1}  &  \textbf{1} &
                             \textbf{1}  &  \textbf{1}  &  \textbf{2}  \\
             &    150     &  \textbf{1}  &  \textbf{1}  &  \textbf{1} &
                             \textbf{1}  &  \textbf{1}  &  \textbf{1}  \\
             &    200     &  \textbf{1}  &  \textbf{1}  &  \textbf{1} &
                             \textbf{1}  &  \textbf{1}  &  \textbf{1}  \\
\hline
\textbf{190} &     50     &  \textbf{1}  &  \textbf{$>$ 9}  &  \textbf{3} &
                             \textbf{1}  &  \textbf{1}  &  \textbf{1}    \\
             &    100     &  \textbf{1}  &  \textbf{1}  &  \textbf{1} &
                             \textbf{1}  &  \textbf{1}  &  \textbf{1}    \\
             &    150     &  \textbf{1}  &  \textbf{1}  &  \textbf{1} &
                             \textbf{1}  &  \textbf{1}  &  \textbf{1}    \\
             &    200     &  \textbf{1}  &  \textbf{1}  &  \textbf{1} &
                             \textbf{1}  &  \textbf{1}  &  \textbf{1}    \\
\hline
\textbf{200} &     50     &  \textbf{1}  &  \textbf{7}  &  \textbf{1}  &
                             \textbf{1}  &  \textbf{1}  &  \textbf{2}    \\
             &    100     &  \textbf{1}  &  \textbf{2}  &  \textbf{1}  &
                             \textbf{1}  &  \textbf{1}  &  \textbf{2}    \\
             &    150     &  \textbf{1}  &  \textbf{1}  &  \textbf{1}  &
                             \textbf{1}  &  \textbf{1}  &  \textbf{1}    \\
             &    200     &  \textbf{1}  &  \textbf{1}  &  \textbf{1}  &
                             \textbf{1}  &  \textbf{1}  &  \textbf{1}     \\
\hline\hline
\end{tabular}
}
\caption{Maxima computed with the CEASD algorithm for three $16^4$, 
$\beta=6.0$ SU(3) configurations.}
\label{ga16}
\end{center}
\end{table}
%==========================================================================
%==========================================================================

In table \ref{ga16} we summarize the performance of the algorithm for the
three gauge configurations considered. For the smallest population, 
the algorithm seems to identify the absolute maximum in 200 generations for 
$N \, \ge \, 180$ (\nsteps = 180). 
For runs up to 50 generations, when 
\nipop = 10 the algorithm sometimes fails the computation of the 
absolute maximum\footnote{
For the remaining 4 configurations,
we verified that for $N$ = 180, 190, 200, \nipop = 10 and for 50
generations the CEASD method computed correctly the absolute 
maximum of $F_U$.}. For runs up to 50 generations and 
$N \, \ge \, 180$, the probability of getting the
absolute maximum is $p_{max} \, = \, 0.67$ for \nipop = 10. The probability of
getting a maximum which is not the absolute maximum in $K$ independent runs
is then $p_{other} \, = \, 0.33^K$, a number which goes rapidly to 
zero\footnote{$p_{other} \, = \, 0.33, ~ 0.11, ~ 0.036, ~ 0.012, ~ 0.004$ for
$K \, = \, 1, ~ 2, ~ 3, ~ 4, ~ 5$.}  with $K$. 
Therefore, it seems reasonable to 
try the use of smaller number of generations, provided multiple 
independent\footnote{For runs up to 50 generations, if one considers the 
results for all the seven configurations $p_{max} \, = \, 0.86$ and 
$p_{other} \, = \, 0.14^K$.} 
runs of the algorithm are done. A possible improvement of the multiple run
situation could be a parallel version of the CEASD algorithm, with the
interchange of chromosomes between the essentially independent populations
every now and then. For the largest population, \nipop = 20, the algorithm
identifies the absolute maximum when 200 generations are considered for
$N \, \ge \, 170$ (\nsteps = 170). 
For runs up to 50 generations, again, the algorithm does not provide the
right maximum. Now, $p_{max} \, = \, 0.75$ for \nipop = 20 and the situation
becomes similar to case discussed previously. Once more, multiple runs
of the CEASD algorithm should be able to identify the absolute maximum of
$F_U$ when using 50 generations. 

In conclusion, if for runs up 50 generations only a multiple independent run
can provide the right answer, when the algorithm uses 200 generations, it
is possible to define \nsteps:
\begin{center}
\begin{tabular}{ccc}
     \nipop &  \mbox{   }  & \nsteps \\
     10   &  &  180 \\
     20   &  &  170.
\end{tabular}
\end{center}

To close this section we report now on the cpu times. On a 
Pentium IV at 2.40 GHz, for a $16^4$ lattice, \nsteps = 200, \nipop = 10 and 
for $\theta \, < \, 10^{-15}$ the cpu time measured required by the CEASD 
algorithm was
\begin{center}
\begin{tabular}{ccc}
     number of generations &  \mbox{   }  & cpu time (s) \\
     50   &  & 112090 \\
     200  &  & 436071
\end{tabular}
\end{center}
meaning that the CEASD algorithm requires about 2211 seconds/generation. For
the same gauge fixing precision, the
steepest descent method requires 1826 seconds. Then, the 
time required by a run with 200 generations is similar to the time
required by 240 multiple steepest descent.
 The CEASD memory 
requirements for the evolutive phase (\npop = 4) are about 236MB.

%=============================================================================
%=============================================================================
\section{Discussion and Conclusions}

In this paper we describe a method for Landau gauge fixing that combines an
evolutionary algorithm with a local optimization method. The ``happy 
marriage'' between the two algorithms is achieved by redefining the
cost function of the EA, in such a way that it becomes an approximation for 
the local maximum in the neighborhood of the chromosome. In order to get the
global maximum, the CEASD algorithm seems to require values for $\theta$ 
of the order of $10^{-7}$ for $8^4$ configurations and $10^{-8}$ for
$16^4$ configurations. Note that the 
CEASD requires only a good approximation of $F_U$
in order to be able to compute the global optimum.

The combined algorithm was tested for three different configurations in two
lattices: a smaller $8^4$ lattice and a larger $16^4$ lattice. For both 
lattices it was possible to identify
a set of parameters for the CEASD method such that,
in a single run, the computed maximum, i.e. the maximum obtained after applying
the steepest descent method to best member of the population of the last 
generation, was always the global maximum
defined from multiple steepest descent runs.

For the smaller lattice the CEASD performed extremely well. Indeed, despite the
relative large number of local maxima, the algorithm seems to be quite
stable in identifying the global maximum of $F_U$ - see 
table \ref{estmilcv2_generations}. 
For the larger lattice, the number of local maxima is much larger when
compared to the $8^4$ lattice. Not only the number of maxima is
larger but they are closer to each other. From the point of view of the global
optimization, this means that the numerical problem in hands is much
harder to solve. Nevertheless, again it was possible to
define a set of parameters
such that the algorithm identified the global maximum in all tested 
configurations - see table \ref{ga16}. Our choice of parameters
for the CEASD algorithm
(200 generations, \nsteps = 100 for $8^4$ lattice and \nsteps = 200
for the larger lattice and for \nipop = 10) is a conservative choice.
As explained before, it is possible to use smaller values of $N$ or 
smaller number of generations. Decreasing $N$  and/or the number of generations
implies decreasing the run time of the CEASD algorithm. However, reducing
$N$ and/or the number of generations should be done with care. Indeed, the
comparative study of the two lattices shows that the complexity of the 
maximisation problem increases with the lattice size, that the method works
better for larger populations, larger values of $N$ and for sufficiently large
number of generations. Nevertheless, the results of the previous section
also show that, for relatively large values of $N$, the probability of
computing the absolute maximum of $F_U$ is large. This suggests that 
a possible solution to the global optimization problem is to perform multiple
independent CEASD runs using lower values of $N$ and/or smaller number of
generations. For sufficient
number of independent runs, in principle, the method should be able to get
the global maximum. A similar situation is found when one relies on
simulated annealing for global optimization problems.

The cpu times required by CEASD algorithm for the two lattice sizes seems to
suggest that the scaling law of the combined method is close to the Fourier 
accelerated SD method, i.e. $V^\delta \, \ln V$ with $\delta$ taking values 
close to 1. A measure of $\delta$ requires necessarily an analysis with more 
lattice sizes\footnote{Assuming a scaling law like $V^\delta \, \ln V$ and
using the cpu times reported in this work, we get 
$\delta_{\mbox{SD}} \, = \, 1.15$ for the Fourier accelerated method and
$\delta_{\mbox{CEASD}} \, = \, 1.24$ for the CEASD algorithm. For a $32^4$
configuration, these numbers mean that a SD gauge fixing requires about
$\tau \, = \, 6 \times 10^4$ s, the CEASD requires 
$72 \times \tau$ seconds to run in 50 
generations and $282 \times \tau$ seconds for a 200 generation run.}. 
This is a numerical intensive problem. We are currently 
engaged in measuring $\delta$ and will report the result elsewhere. 
Na\"ively, one expects that gauge fixing to the minimal Landau 
gauge with CEASD is as demanding as performing a gauge fixing with the SD 
method. 

In principle, it is possible to combine the EA with any local optimisation
method. Faster local methods will produce faster combined algorithms. 
The time required by a combined algorithm is strongly dependent on
the performance of the local method. The gauge fixing is a computational
intense problem. Therefore, it is important to investigate new and more
performant local methods.

The CEASD algorithm described here for Landau gauge fixing seems to solve
the problem of the minimal Landau gauge fixing. Moreover, the method
is suitable to be used with other gauge conditions that also suffer from the 
Gribov ambiguity and are currently used in lattice gauge theory. 
The effects of Gribov copies in QCD correlation functions remains to be
investigated \cite{nosso}.

P. J. S. acknowledges financial support from the portuguese FCT. This
work was in part supported by fellowship Praxis/P/FIS/14195/98 under 
project Optimization in Physics and in part from grant  SFRH/BD/10740/2002.

\end{document}